\documentclass[journal]{IEEEtran1}
\usepackage{epsfig}
\usepackage{subfigure}
\usepackage{graphicx}
\usepackage{times}
\usepackage{cite}
\usepackage{enumerate}
\usepackage{amssymb}
\usepackage{algorithm}
\usepackage{algorithmic}
\usepackage{multirow}
\usepackage{amsmath}
\usepackage{braket}
\usepackage{multicol}

\newtheorem{theorem}{Theorem}[section]

\begin{document}

\title{Optimal Utilization of a Cognitive Shared Channel with a Rechargeable Primary Source Node}

\author{
\authorblockN{Nikolaos Pappas, Jeongho Jeon, Anthony Ephremides, and Apostolos Traganitis}
\thanks{The material in this paper is presented in part at 2011 IEEE Information Theory Workshop.}
\thanks{Nikolaos Pappas and Apostolos Traganitis are with the Computer Science Department, University of Crete Greece and the Institute of Computer Science, Foundation for Research and Technology - Hellas (FORTH) (e-mail: \{npapas, tragani\}@ics.forth.gr).}
\thanks{Jeongho Jeon and Anthony Ephremides are with the Department of Electrical and Computer Engineering and the Institute for Systems Research, University of Maryland, College Park, MD 20742
USA (e-mail: \{jeongho, etony\}@umd.edu).}}

\maketitle

\begin{abstract}

This paper considers the scenario in which a set of nodes share a common channel. Some nodes have a rechargeable battery and the others are plugged to a reliable power supply and, thus, have no energy limitations. We consider two source-destination pairs and apply the concept of cognitive radio communication in sharing the common channel. Specifically, we give high-priority to the energy-constrained source-destination pair, i.e., primary pair, and low-priority to the pair which is free from such constraint, i.e., secondary pair. In contrast to the traditional notion of cognitive radio, in which the secondary transmitter is required to relinquish the channel as soon as the primary is detected, the secondary transmitter not only utilizes the idle slots of primary pair but also transmits along with the primary transmitter with probability $p$. This is possible because we consider the general multi-packet reception model. Given the requirement on the primary pair's throughput, the probability $p$ is chosen to maximize the secondary pair's throughput. To this end, we obtain two-dimensional maximum stable throughput region which describes the theoretical limit on rates that we can push into the network while maintaining the queues in the network to be stable. The result is obtained for both cases in which the capacity of the battery at the primary node is infinite and also finite.


\end{abstract}

\begin{keywords}
cognitive network, stochastic energy harvesting, stability analysis, multipacket reception capability
\end{keywords}

\section{Introduction}
\label{sec:intro}


\PARstart{C}{ognitive} radio communication provides an efficient means of sharing radio spectrum between users having different priority \cite{zhao:survey}. The high-priority user, called primary, is allowed to access the channel whenever it needs, while the low-priority user, called secondary, is required to make a decision on its transmission based on what the primary user does. The system considered in this paper is comprised of nodes that are either subject to energy availability constraint imposed by the battery status and stochastic recharging process or free from such constraint by assuming that they are connected to a constant power source.

\begin{figure}[t]
\centering
\epsfig{file=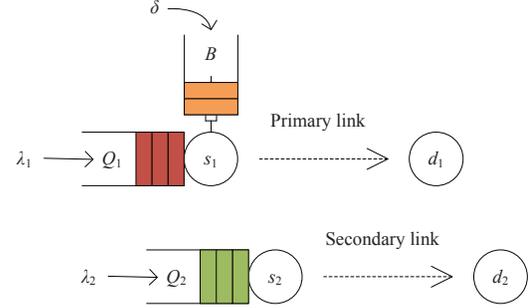,angle=0,width=0.38\textwidth}
\caption{An example cognitive communication system}
\label{fig:model}
\end{figure}

In this paper, we consider the simple cognitive system of two source-destination pairs as shown in Fig.~\ref{fig:model} and derive the \emph{maximum stable throughput region}
for a cognitive access protocol on the general multipacket reception channel model\footnote{When compared to collision channel model, it better captures the effects of fading, attenuation and interference at the physical layer.} in which a transmission may succeed even in the presence of interference \cite{ghez:stability, tong:multipacket, naware:stability}. The secondary node can take advantage of such an additional reception capability by transmitting simultaneously with the primary. We adopt a similar cognitive access protocol proposed in \cite{rong:cooperation}, and also studied in \cite{kompella:stable}, in which the secondary node not only utilizes the idle periods of the primary node, but also competes with the primary by randomly accessing the channel to increase its own throughput. However, the secondary user is still required to coordinate its transmission in order not to hamper the required throughput level of the primary link given the energy harvesting rate and this is done by appropriately choosing the random access probability.

To position our contribution with respect to the recent literature, we start a brief background review. In \cite{ozel:information}, the capacity of the additive white Gaussian noise channel with stochastic energy harvesting at the source was shown to be equal to the capacity with an average power constraint given by the energy harvesting rate. However, like most of information-theoretic research, the result is obtained for point-to-point communication with an always backlogged source. In \cite{jeon:stability}, the slotted ALOHA protocol was considered for a network of nodes having energy harvesting capability and the maximum stable throughput region was obtained for bursty traffic. An exact characterization of the region was given in the paper for a two-node case over a collision channel. The analysis is not trivial even for such a simple network because the service process of a node not only depends on the status of its battery but also on the idleness or not of the other node. Note that the reason why the exact region is known only for the two-node and the three-node cases (even without energy availability constraints) is the \emph{interaction} between the queues of the nodes \cite{tsybakov:ergodicity, rao:stability, Szpankowski:stability, luo:stability}.

The initial study of a simple model involving only two source-destination pairs is not only instructive but also necessary. The reason is that the interaction between nodes causes considerable difficulties at the analytical level, and yet, reveals major insights at the conceptual level.
In addition, we use the stochastic dominance technique and Loynes' theorem \cite{loynes:stability} for the stability of stationary system to solve the problem. Also, as pointed out in \cite{jeon:stability}, it is important to note that the "service process" of the battery, i.e., the use of its energy, is independent of whether the transmission is successful or not.

The rest of the paper is organized as follows. In Section~\ref{sec:model}, we define the stability region, describe the channel model, and explain the packet arrival and energy harvesting models. In Section~\ref{sec:cognitive}, we present the conditions for stability of the considered cognitive access protocol when the capacity of the battery at the primary node is assumed to be infinite. The proof of the result is given in Section~\ref{sec:analysis} which utilizes the stochastic dominance technique and arguments similar to those used in \cite{jeon:stability} and \cite{rao:stability}. In Section~\ref{sec:finite}, we extend the result to the case when the capacity of battery is finite. As will be shown, the stability region for the case with finite capacity battery is a subset of that for the case with infinite capacity battery. For comparison's sake, in Section~\ref{sec:collision}, the result obtained in Section~\ref{sec:cognitive} is derived again for the case without multipacket reception capability, i.e., for a collision channel with additional probabilistic erasures. Finally, we draw some conclusions in Section~\ref{sec:conclusion}.

\section{System Model}\label{sec:model}

We consider a time-slotted communication system consisting of two primary and secondary source-destination pairs of nodes, $(s_1, d_1)$ and $(s_2, d_2)$, respectively, as shown in Fig. \ref{fig:model}. Each source node has an infinite capacity buffer $Q_i$ $(i \in \set{1,2})$ for storing arriving packets of fixed length. The secondary node is plugged to a reliable power supply, whereas the primary node is powered through a random time-varying renewable energy process and has a battery $B$ for storing energy which is assumed to be harvested in a certain unit from the environments. The capacity of the battery is denoted by $c$. We first consider the case with $c=\infty$ and, after that, we relax $c$ to take any finite integer value. The slot duration is equal to the transmission time of a single packet and one unit of energy is consumed in each transmission. The packet arrival and energy harvesting processes are all modeled as independent Bernoulli processes of rate $\lambda_i$ and $\delta$ per slot, respectively. The primary node is considered \emph{active} if both $Q_1$ and $B_1$ are nonempty at the same time. Similarly, the secondary is called \emph{active} if $Q_2$ is nonempty. Otherwise, they are called \emph{idle}.

A shared channel is assumed and a transmission is said to be successful if the received signal-to-interference-plus-noise-ratio (SINR) exceeds a certain threshold which depends on the modulation scheme, the target bit-error-rate, and the number of bits in the packet (i.e., the transmission rate for a fixed packet duration). Denote by $q_{i/ \boldsymbol{I}}$ the probability that the transmission by source $i$ succeeds given that the sources in $\boldsymbol{I}$ are transmitting simultaneously. Specifically, in our cognitive communication system in Fig. \ref{fig:model}, the following success probabilities are of interest:
\begin{equation*}
    q_{1/1}, \ q_{2/2}, \ q_{1/1,2}, \ q_{2/1,2}
\end{equation*}
and it is assumed that $q_{1/1} \geq q_{1/1,2}$ and $q_{2/2} \geq q_{2/1,2}$. Define $\Delta_1 = q_{1/1} - q_{1/1,2}$ and $\Delta_2 = q_{2/2} - q_{2/1,2}$. In case that the simultaneous transmissions always fail, we have $q_{i/1,2} = 0$ for all $i$.

Denote by $Q_i^t$ the length of $Q_i$ at the beginning of time slot $t$, the queue is said to be \emph{stable} if
\begin{equation}\label{eqn:definition_stability}
    \lim_{ {x} \rightarrow \infty} \lim_{t \rightarrow \infty} \mathrm{Pr}[Q_i^t < {x}] = 1
\end{equation}
Loynes' theorem \cite{loynes:stability} states that if the arrival and service processes of a queue are strictly jointly stationary and the average arrival rate is less than the average service rate, then the queue is stable. If the average arrival rate is greater than the average service rate, then the queue is unstable and the value of $Q_i^t$ approaches to infinity almost surely. The stability region of the system is defined as the set of arrival rate vectors $\boldsymbol{\lambda}=(\lambda_1, \lambda_2)$ for which the queues in the system are stable.

\section{Main Results}\label{sec:cognitive}

This section describes the cognitive access protocol and presents our main results concerning its stability. The proofs of the results are presented in the next section.

\subsection{Description of the cognitive access protocol}\label{sec:sec:protocol}

The opportunistic cognitive access protocol proposed in \cite{rong:cooperation} and also used in \cite{kompella:stable} is modified and studied again in the context of the energy harvesting environment. The energy-constrained primary node $s_1$ (see Fig. \ref{fig:model}) transmits a packet whenever it is active. Note that the transmission by the primary node $s_1$ is independent of the secondary node $s_2$. On the other hand, the transmission by the secondary node $s_2$ must be chosen in a careful manner in order not to impede the primary's performance guarantees. Under our cognitive access protocol, node $s_2$ observes the status of $s_1$ and if $s_1$ is idle, i.e., either $Q_1$ or $B_1$ is empty, it transmits with probability $1$ if its own packet queue $Q_2$ is nonempty. Otherwise, if $s_1$ is active, $s_2$ transmits with probability $p$ to take advantage of the multipacket reception capability by transmitting along with the primary node although at the same time it risks impeding the primary node's success. The design objective is to choose the transmission probability $p$ such that the secondary's throughput is maximized while maintaining the stability of primary's packet queue at given packet arrival and energy harvesting rates.

\subsection{Stability Criteria}\label{sec:sec:main}

\begin{figure*}
\begin{equation}\label{eqn:main_a}
    \mathcal{R}_1'=\left\{ (\lambda_{1},\lambda_{2}): \frac{\Delta_2}{q_{1/1,2}q_{2/2}}\lambda_{1}+\frac{\lambda_{2}}{q_{2/2}} < 1, \ \ 0 \leq \lambda_{1} \leq \delta q_{1/1,2} \right\}
\end{equation}
\end{figure*}
\begin{figure*}
\begin{equation}\label{eqn:main_b}
    \mathcal{R}_1'' \! = \! \left\{ (\lambda_{1},\lambda_{2}): \frac{q_{2/1,2} \lambda_1 + \Delta_1 \lambda_2}{\delta q_{1/1} q_{2/1,2} + \Delta_1 q_{2/2} (1-\delta)} < 1, \ \ \delta q_{1/1,2} < \lambda_{1} < \delta q_{1/1} \right\}
\end{equation}
\end{figure*}
\begin{figure*}
\begin{equation}\label{eqn:main_e}
\mathcal{R}_{2}''=\left\{ (\lambda_{1},\lambda_{2}): \frac{q_{2/1,2}\lambda_{1}+\Delta_1 \lambda_{2}}{\delta q_{1/1} q_{2/1,2} +\Delta_1 q_{2/2} (1-\delta)} < 1, \ \ (1-\delta)q_{2/2} < \lambda_{2} < (1-\delta)q_{2/2} + \delta q_{2/1,2} \right\}
\end{equation}
\end{figure*}

\begin{figure}[t]
\centering \subfigure[The case with $\eta > 0$]
{\label{fig:R1_eta_g_0}\epsfig{file=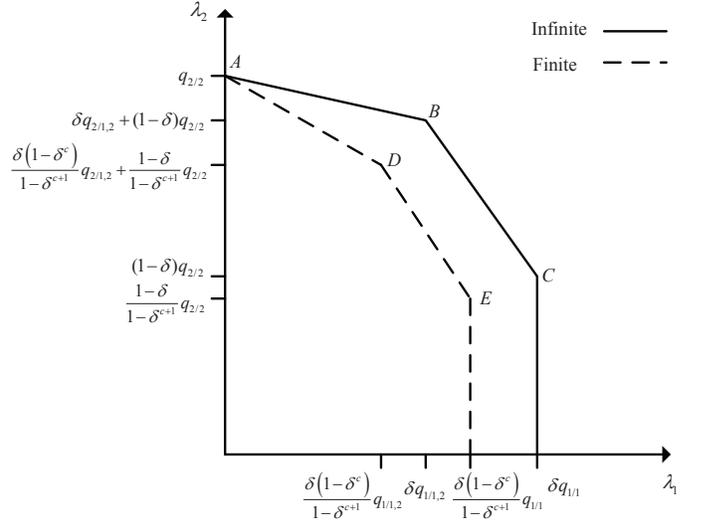,angle=0,width=0.49\textwidth}}
\hspace{0.3cm}\centering \subfigure[The case with $\eta \leq 0$]
{\label{fig:R1_eta_leq}\epsfig{file=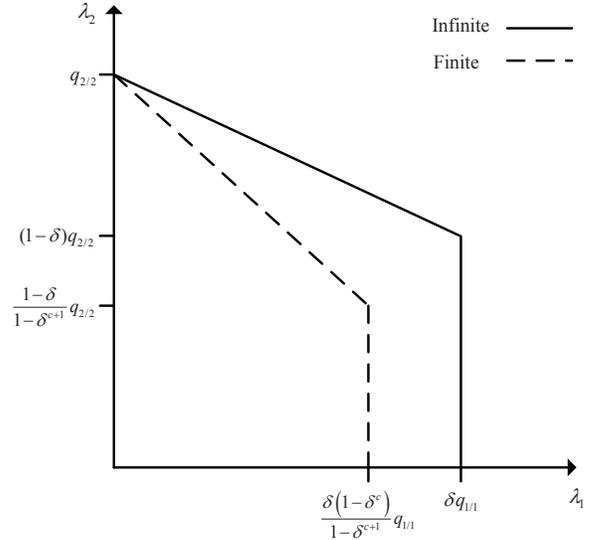,angle=0,width=0.42\textwidth}}
\caption{The subregion $\mathcal{R}_1$ with multipacket reception capability (solid and dotted lines depict the case when the capacity of the primary node is infinite and finite, respectively.)}
\label{fig:R1}
\end{figure}

\begin{figure}[t]
\centering
\epsfig{file=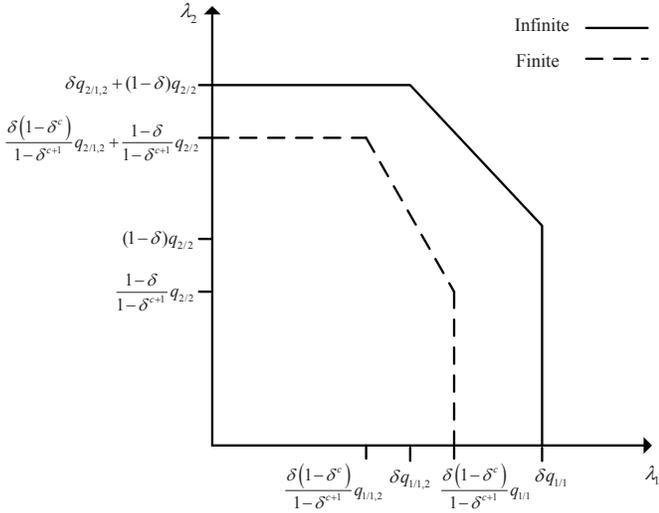,angle=0,width=0.48\textwidth}
\caption{The subregion $\mathcal{R}_2$ with multipacket reception capability (solid and dotted lines depict the case when the capacity of the primary node is infinite and finite, respectively.)}
\label{fig:R2}
\end{figure}

Denote by $\mathcal{R}$ the stability region of the system by considering all possible values of $p$ and define $\eta=q_{1/1}q_{2/1,2}+q_{2/2}q_{1/1,2}-q_{2/2}q_{1/1}$. Note that $\eta$ reflects the degree of multipacket reception capability. In the case of a collision channel in which $q_{1/1}=q_{2/2}=1$ and $q_{1/1,2}=q_{2/1,2}=0$, $\eta=-1$. It is clear that $\eta$ increases as the multipacket reception capability improves.
\begin{theorem}\label{thm:main_infinite_mpr}
The stability region of the cognitive multiaccess system is described by
\begin{equation}
    \mathcal{R} = \mathcal{R}_1 \bigcup \mathcal{R}_2
\end{equation}
where the subregion $\mathcal{R}_1$ is described as follows:
\begin{itemize}
\item If $\eta>0$, $\mathcal{R}_1 = \mathcal{R}_1' \bigcup \mathcal{R}_1''$ where $\mathcal{R}_1'$ and $\mathcal{R}_1''$ are given by \eqref{eqn:main_a} and \eqref{eqn:main_b}.

\item If $\eta \leq 0$,
\begin{equation}\label{eqn:main_c}
    \mathcal{R}_1=\left\{ (\lambda_{1},\lambda_{2}): \frac{\lambda_1}{q_{1/1}}+\frac{\lambda_{2}}{q_{2/2}} < 1, \ \ \lambda_{1} < \delta q_{1/1} \right\}
\end{equation}
\end{itemize}
and the subregion $\mathcal{R}_2$ is described as $\mathcal{R}_2 = \mathcal{R}_2' \bigcup \mathcal{R}_2''$ with
\begin{equation}\label{eqn:main_d}
\mathcal{R}_{2}'=\left\{ (\lambda_{1},\lambda_{2}): \lambda_{1} < \delta q_{1/1}, 0 \leq \lambda_{2} \leq (1-\delta) q_{2/2}  \right\}
\end{equation}
and $\mathcal{R}_2''$ as given by \eqref{eqn:main_e}.
\begin{proof}
The proof is given in Section \ref{sec:analysis}.
\end{proof}
\end{theorem}
The optimal $p^*$ achieving the boundary of the stability region is explicitly given in the following section. The subregion $\mathcal{R}_1$ is depicted in Fig.~\ref{fig:R1} with solid line. Specifically, if $\eta > 0$, the line segments $AB$ and $BC$ correspond to the boundaries due to the inequalities \eqref{eqn:main_a} and \eqref{eqn:main_b}, respectively. The subregion $\mathcal{R}_2$ is also illustrated in the Fig.~\ref{fig:R2} with solid line. Note that when $\eta > 0$, $\mathcal{R}_2$ is always contained in $\mathcal{R}_1$, i.e., $\mathcal{R}_2 \subset \mathcal{R}_1$, which is not necessarily true if $\eta \leq 0$.

\section{Analysis using Stochastic Dominance}\label{sec:analysis}
\label{sec:analysis}

Under the cognitive access protocol described in Section \ref{sec:sec:protocol}, the expressions for the  average service rates seen by $s_1$ and $s_2$ are given by
\begin{multline}\label{eqn:service_rate_primary}
\mu_1= q_{1/1} \mathrm{Pr}[B_1 \neq 0, Q_2 = 0] + q_{1/1,2} \mathrm{Pr}[B_1 \neq 0, Q_2 \neq 0] p \\ + q_{1/1} \mathrm{Pr}[B_1 \neq 0, Q_2 \neq 0] (1-p)
\end{multline}
and
\begin{multline}\label{eqn:service_rate_secondary}
\mu_2 = q_{2/2} (1- \mathrm{Pr}[B_1 \neq 0, Q_1 \neq 0] )\\ + q_{2/1,2} \mathrm{Pr}[B_1 \neq 0, Q_1 \neq 0] p
\end{multline}
%
Note that computing the average service rates $\mu_{1}$ and $\mu_{2}$ requires the specifications of a joint probability of doublets $(B_1, Q_2)$ and $(B_1, Q_1)$, respectively. Since, however, $Q_1$, $Q_2$, and $B_1$ are all \emph{interacting}, it is difficult to track them. We bypass this difficulty by utilizing the idea of stochastic dominance \cite{rao:stability}. That is we first construct parallel dominant systems in which one of the nodes transmits dummy packets even when its packet queue is empty. The essence of the dominant system is to make the analysis tractable by decoupling the interaction between the queues. Since the queue sizes in the dominant system are, at all times, at least as large as those of the original system, the stability region of the dominant system inner bounds that of the original system. It turns out however that the stability region obtained using this stochastic dominance technique coincides with that of the original system which will be discussed in detail later in this section. Thus, the stability regions for both the original and the dominant systems are the same.

\subsection{The first dominant system: secondary node transmits dummy packets}

Construct a hypothetical system in which the secondary node $s_2$ transmits dummy packets when its packet queue is empty. Hence $s_2$ transmits with probability 1 whenever $s_1$ is idle and with probability $p$ if $s_1$ is active. As a result, the average service rate of $s_1$ in \eqref{eqn:service_rate_primary} reduces to
\begin{equation}\label{eqn:service_rate_primary_d1}
\mu_1= q_{1/1,2} \mathrm{Pr}[B_1 \neq 0] p + q_{1/1} \mathrm{Pr}[B_1 \neq 0] (1-p)
\end{equation}
Since $s_1$ transmits with probability 1 whenever it is active, if $Q_1$ is saturated\footnote{Note that in describing the service rates in \eqref{eqn:service_rate_primary} and \eqref{eqn:service_rate_secondary}, it is assumed that the corresponding packet queue is nonempty. This is simply because if the queue is empty, the "server" becomes idle.}, $B_1$ is modeled as a decoupled discrete-time $M/D/1$ system with arrival and service rates $\delta$ and $1$, respectively. It follows from Little's theorem that $B_1$ is nonempty for a fraction of time $\delta$ \cite{kleinrock:queueing}. Consequently, we have
\begin{equation}\label{eqn:stability_primary}
\mu_1= \delta( q_{1/1,2} p + q_{1/1} (1-p))
\end{equation}

For $\lambda_1$ satisfying $\lambda_1 < \mu_1$, i.e., when $Q_1$ in this dominant system is stable, we now obtain the average service rate of $s_2$. We note from \eqref{eqn:service_rate_secondary} that the probability of $s_1$ being active, i.e., $\mathrm{Pr}[B_1 \neq 0, Q_1 \neq 0]$, needs to be specified beforehand. For this, we take an approach similar to the one used in \cite{jeon:stability}. The approach utilizes a simple property of a stable system, that is the rate of what comes is equal to the rate of what goes out. Given the fact that $s_1$ is active, the average number of packets out of $Q_1$ is given by $q_{1/1,2} p + q_{1/1} (1-p)$. Because the average number of packets into $Q_1$ is $\lambda_1$ and, because it satisfies $\lambda_1 < \mu_1$, the fraction of active slots must be
\begin{equation}
\mathrm{Pr}[B_1 \neq 0, Q_1 \neq 0]= \frac{\lambda_1}{q_{1/1,2} p + q_{1/1} (1-p)}
\end{equation}
After some manipulation, the average service rate of $s_2$ can be obtained from \eqref{eqn:service_rate_secondary} as
\begin{equation}\label{eqn:service_rate_secondary_d1}
\mu_2 = \frac{q_{2/1,2}p - q_{2/2}}{q_{1/1,2}p + q_{1/1}(1-p)} \lambda_1 + q_{2/2}
\end{equation}
By applying Loynes' theorem, we find that the stability condition for the dominant system is given by
\begin{equation}\label{eqn:stability_region_d1_secondary}
\lambda_2 < \frac{q_{2/1,2}p - q_{2/2}}{q_{1/1,2}p + q_{1/1}(1-p)} \lambda_1 + q_{2/2}
\end{equation}
when
\begin{equation}\label{eqn:stability_region_d1_primary}
\lambda_1 < \delta( q_{1/1,2} p + q_{1/1} (1-p))
\end{equation}

An important observation made in \cite{rao:stability} is that the stability conditions obtained by using stochastic dominance technique are not merely sufficient conditions for the stability of the original system but are sufficient and necessary conditions. The \emph{indistinguishability} argument applies to our problem as well. Based on the construction of the dominant system, it is easy to see that the queues of the dominant system are always larger in size than those of the original system, provided they are both initialized to the same value.
%
Therefore, given $\lambda_{1}<\mu_{1}$, if for some $\lambda_{2}$, the queue at $s_{2}$ is stable in the dominant system then the corresponding queue in the original system must be stable; conversely, if for some $\lambda_{2}$ in the dominant system, the node $s_{2}$ saturates, then it will not transmit dummy packets, and as long as $s_{2}$ has a packet to transmit, the behavior of the dominant system is identical to that of the original system because the action of dummy packet transmissions is employed increasingly rarely as we approach the stability boundary. Therefore, we can conclude that the original system and the dominant system are indistinguishable at the boundary points.

The portion of the stable throughput region by the first dominant system is given by the closure of the rate pairs $(\lambda_{1},\lambda_{2})$ described by \eqref{eqn:stability_region_d1_secondary} and \eqref{eqn:stability_region_d1_primary} as $p$ varies over $[0,1]$. To obtain the closure of the rate pair, we first fix $\lambda_{1}$ and maximize $\lambda_{2}$ as $p$ varies over $[0,1]$. By replacing $\lambda_{1}$ by $x$ and $\lambda_{2}$ by $y$, the boundary of the stability region for fixed $p$ can now be written as
\begin{equation}\label{eqn:optimization_obj}
y = \frac{q_{2/1,2}p - q_{2/2}}{q_{1/1,2}p + q_{1/1}(1-p)} x + q_{2/2}
\end{equation}
for $0\leq x \leq \delta( q_{1/1,2} p + q_{1/1} (1-p))$. Differentiating $y$ with respect to $p$ yields,
\begin{equation}
\frac{dy}{dp}=\frac{\eta x}{\left(q_{1/1}+p(q_{1/1,2}-q_{1/1}) \right)^{2}}
\end{equation}
where $\eta$ is defined in Section \ref{sec:sec:main}. It can be observed that the denominator is strictly positive and the numerator can be positive or negative depending on the value of $\eta$.
\begin{itemize}

\item If $\eta > 0$, the first derivative is strictly positive, and $y$ is an increasing function of $p$. Therefore $p^{*}=1$. However, this is valid only if $0\leq x \leq \delta( q_{1/1,2} p + q_{1/1} (1-p))$. Thus, $p^{*}$ can take a value of 1 only if $0\leq x \leq \delta q_{1/1,2}$.  Substituting $p^{*}=1$ into \eqref{eqn:optimization_obj} gives the boundary of the subregion characterized by \eqref{eqn:main_a}.

\item If $\eta > 0$ and $x> \delta q_{1/1,2}$, then $p^{*}=\frac{\delta q_{1/1} -x }{\delta (q_{1/1} - q_{1/1,2})}$. By substituting $p^{*}$ into \eqref{eqn:optimization_obj} and after some simple algebra, we obtain the boundary of the subregion characterized by \eqref{eqn:main_b}.


\item If $\eta \leq 0$, the derivative is non-positive for all feasible $p$ and, thus, $y$ is a decreasing function of $p$ in the range of all possible values of $x$. Therefore, $p^{*}=0$ and the stability region is given in \eqref{eqn:main_c}.

\end{itemize}
Note that for the first dominant system the value of $\lambda_{1}$ is upper bounded by the term $\delta q_{1/1}$.

\subsection{The second dominant system: primary node transmits dummy packets}

In the previous section, we obtained the stability region of the first dominant system which yields one part of the stability region of the original system. To finalize the analysis, consider the complementary dominant system in which the primary node $s_1$ transmits dummy packets whenever its packet queue is empty, and the secondary node $s_2$ behaves exactly as in the original system. Even in the dominant system, however, $s_1$ cannot transmit if its battery is empty. Therefore, the average service rate of $s_2$ in \eqref{eqn:service_rate_secondary} reduces to
\begin{equation} \label{eqn:service_rate_secondary_d2}
\mu_{2}= q_{2/2} \left(1-\mathrm{Pr}[B_{1} \neq 0]\right) + q_{2/1,2} \mathrm{Pr}[B_{1} \neq 0]p
\end{equation}
Since $s_1$ transmits with probability 1 whenever its battery is nonempty, $B_1$ is modeled as a decoupled discrete-time $M/D/1$ system with arrival rate $\delta$ and service rate 1. Consequently, \eqref{eqn:service_rate_secondary_d2} becomes
\begin{equation} \label{eqn:service_rate_secondary_d2_1}
\mu_{2}= q_{2/2} \left(1- \delta \right) + q_{2/1,2} \delta p
\end{equation}
From Little's theorem, the probability that $Q_2$ is nonempty for some $\lambda_2 < \mu_2$ is given by
\begin{equation}\label{eqn:prob_q2_nonempty_d2}
\mathrm{Pr}[Q_{2} \neq 0]=\frac{\lambda_{2}}{q_{2/2} \left(1- \delta \right) + q_{2/1,2} \delta p}
\end{equation}

Because in this dominant system $B_1$ is decoupled, i.e., independent, from the rest of the system, we can rewrite the average service rate of $s_1$ in \eqref{eqn:service_rate_primary} as
\begin{multline} \label{eqn:service_rate_primary_d2}
\mu_{1}=\mathrm{Pr}[B_{1} \neq 0] \{q_{1/1}\left(1-\mathrm{Pr}[Q_{2} \neq 0] \right) \\
+q_{1/1,2} \mathrm{Pr}[Q_{2} \neq 0]p + q_{1/1} \mathrm{Pr}[Q_{2} \neq 0](1-p)\}
\end{multline}
Plugging \eqref{eqn:prob_q2_nonempty_d2} into \eqref{eqn:service_rate_primary_d2} and, after some manipulations, we find the stability condition for this dominant system is given by
%

\begin{equation}\label{eqn:stability_region_d2_primary}
\lambda_{1} < \mu_{1}=  \frac{\delta p (q_{1/1,2} - q_{1/1})}{(1-\delta)q_{2/2} + \delta p q_{2/1,2}} \lambda_2 + \delta q_{1/1}
\end{equation}
for
\begin{equation}\label{eqn:stability_region_d2_secondary}
\lambda_{2} <  \left(1- \delta \right) q_{2/2} + \delta p q_{2/1,2}
\end{equation}
The indistinguishability argument at saturations holds here as well.

To specify the boundary of the stability region which is the closure of the rate pairs $(\lambda_1, \lambda_2)$ over feasible $p$, we follow the same methodology as in the previous section. By replacing $\lambda_1$ and $\lambda_2$ by $y$ and $x$, respectively, the boundary for fixed $p$ is written as
\begin{equation} \label{eqn:optimization_obj_2}
y=\frac{\delta p (q_{1/1,2} - q_{1/1})}{(1-\delta)q_{2/2} + \delta p q_{2/1,2}} x + \delta q_{1/1}
\end{equation}
for $0 \leq x \leq \left(1- \delta \right) q_{2/2} + \delta p q_{2/1,2} $. It is not difficult to see that its first derivative with respect to $p$ is given as
\begin{equation}
\frac{dy}{dp}=- \frac{\theta x}{\left((1-\delta)q_{2/2} + \delta p q_{2/1,2} \right)^{2}}
\end{equation}
where $\theta = \delta (1 - \delta) q_{2/2} (q_{1/1} - q_{1/1,2})$. Since $\theta$ is always non-positive under our assumption, $y$ is a non-increasing function of $p$. Therefore, the optimal value of $p^{*}$ maximizing $y$ is 0 but this is valid only if the condition $0 \leq x \leq \left(1- \delta \right) q_{2/2} + \delta p q_{2/1,2} $ is met. At $p=0$, it becomes $0 \leq x \leq \left(1- \delta \right) q_{2/2}$. Substituting $p^* = 0$ into \eqref{eqn:optimization_obj_2} yields \eqref{eqn:main_d}.
If $x> \left(1- \delta \right) q_{2/2}$, we obtain $p^{*}=\frac{x - (1-\delta)q_{2/2} }{\delta q_{2/1,2}}$. By substituting $p^*$ into \eqref{eqn:optimization_obj_2}, we obtain \eqref{eqn:main_e}.
Note that in obtaining the stability region for this dominant system, it is assumed that $\lambda_2 < \mu_2$. At $\lambda_1 = 0$, the optimal transmission probability of the secondary node is $p=1$ which gives the upper bound on $\lambda_2$ in \eqref{eqn:main_e}.

\section{The Case with Finite Capacity Battery}\label{sec:finite}

We now consider a realistic scenario in which the primary node is equipped with a battery whose capacity is finite. The harvested energy units can be stored only if the battery is not fully charged.

\begin{figure*}
\begin{equation}\label{eqn:main_a_f}
    \mathcal{R}_1'=\left\{ (\lambda_{1},\lambda_{2}): \frac{\Delta_2}{q_{1/1,2}q_{2/2}}\lambda_{1}+\frac{\lambda_{2}}{q_{2/2}} < 1, \ \ 0 \leq \lambda_{1} \leq \frac{\delta (1-\delta^{c})}{1-\delta^{c+1}} q_{1/1,2} \right\}
\end{equation}
\end{figure*}
\begin{figure*}
\begin{equation}\label{eqn:main_b_f}
    \mathcal{R}_1'' \! = \! \left\{ (\lambda_{1},\lambda_{2}): \frac{q_{2/1,2} \lambda_1 + \Delta_1 \lambda_2}{\frac{\delta(1-\delta^c)}{1-\delta^{c+1}} q_{1/1} q_{2/1,2} + \Delta_1 q_{2/2} \frac{ 1-\delta}{1-\delta^{c+1}}} < 1, \ \ \frac{\delta(1-\delta^c)}{1-\delta^{c+1}} q_{1/1,2} < \lambda_{1} < \frac{\delta(1-\delta^c)}{1-\delta^{c+1}}  q_{1/1} \right\}
\end{equation}
\end{figure*}
\begin{figure*}
\begin{equation}\label{eqn:main_e_f}
\mathcal{R}_{2}''=\left\{ (\lambda_{1},\lambda_{2}): \frac{q_{2/1,2}\lambda_{1}+\Delta_1 \lambda_{2}}{\frac{\delta(1-\delta^c)}{1-\delta^{c+1}} q_{1/1} q_{2/1,2} +\Delta_1 q_{2/2} \frac{1-\delta}{1-\delta^{c+1}}} < 1, \ \ \frac{1-\delta}{1-\delta^{c+1}} q_{2/2} < \lambda_{2} < \frac{1-\delta}{1-\delta^{c+1}}q_{2/2} + \frac{\delta(1-\delta^c)}{1-\delta^{c+1}} q_{2/1,2} \right\}
\end{equation}
\end{figure*}

\begin{theorem}
The stability region of the cognitive multiaccess system with finite battery is described by
\begin{equation}
    \mathcal{R} = \mathcal{R}_1 \bigcup \mathcal{R}_2
\end{equation}
where the subregion $\mathcal{R}_1$ is described as follows:
\begin{itemize}
\item If $\eta>0$, $\mathcal{R}_1 = \mathcal{R}_1' \bigcup \mathcal{R}_1''$ where $\mathcal{R}_1'$ and $\mathcal{R}_1''$ are given by \eqref{eqn:main_a_f} and \eqref{eqn:main_b_f}. The optimal probabilities $p*$ achieving the boundaries of the subregions $\mathcal{R}_1'$ and $\mathcal{R}_1''$ are obtained as $p^*=1$ and $p^*= \left(\frac{\delta(1-\delta^c)}{1-\delta^{c+1}}q_{1/1}-\lambda_1 \right) / \left(\frac{\delta(1-\delta^c)}{1-\delta^{c+1}}\Delta_1\right)$, respectively.

\item If $\eta \leq 0$,
\begin{equation}\label{eqn:main_c_f}
    \mathcal{R}_1=\left\{ (\lambda_{1},\lambda_{2}): \frac{\lambda_1}{q_{1/1}}+\frac{\lambda_{2}}{q_{2/2}} < 1, \ \ \lambda_{1} < \frac{\delta(1-\delta^c)}{1-\delta^{c+1}} q_{1/1} \right\}
\end{equation}
The optimal $p*$ achieving the boundary is zero.
\end{itemize}
The subregion $\mathcal{R}_2$ is described as $\mathcal{R}_2 = \mathcal{R}_2' \bigcup \mathcal{R}_2''$ with
\begin{equation}\label{eqn:main_d_f}
\mathcal{R}_{2}'=\left\{ (\lambda_{1},\lambda_{2}): \lambda_{1} < \frac{\delta (1-\delta^c)}{1-\delta^{c+1}} q_{1/1}, 0 \leq \lambda_{2} \leq \frac{1-\delta}{1-\delta^{c+1}} q_{2/2}  \right\}
\end{equation}
and $\mathcal{R}_2''$ as given by \eqref{eqn:main_e_f}. The optimal $p^*$ achieving the boundaries of the subregions $\mathcal{R}_2'$ and $\mathcal{R}_2''$ are obtained as $p^*=0$ and $p^*= \left( \lambda_2 - q_{2/2} \frac{1-\delta}{1-\delta^{c+1}} \right) / \left( \frac{\delta(1-\delta^c)}{1-\delta^{c+1}}q_{2/1,2} \right)$, respectively.
\begin{proof}
In the dominant system in which the primary node transmits dummy packets when its queue is empty, $B$ is decoupled from the remaining of the system and modeled as a discrete-time $M/D/1/c$ system with arrival and service rates $\delta$ and $1$, respectively. We know in that case that $B$ is always ergodic and nonempty with
\begin{equation}
\mathrm{Pr}[B \neq 0]=\frac{\delta(1-\delta^c)}{1-\delta^{c+1}}
\end{equation}
with $\delta$ strictly less than $1$. If $\delta = 1$, $B$ is nonempty with probability 1 which is not of our interest since we can rule out the role of the battery in the steady-state. The rest of the proof is similar to that of Theorem~\ref{thm:main_infinite_mpr}.

\end{proof}
\end{theorem}

The subregion $\mathcal{R}_1$ is depicted in Fig.~\ref{fig:R1} with dotted line. Specifically, if $\eta > 0$, the line segments $AD$ and $DE$ correspond to the boundaries due to the inequalities \eqref{eqn:main_a_f} and \eqref{eqn:main_b_f}, respectively. The subregion $\mathcal{R}_2$ is also plotted in Fig.~\ref{fig:R2} with dotted line. One can easily observe from the figures that the stability region for the case with finite capacity battery is always a subset of that for the case with infinite capacity battery. Also, note that as $c \rightarrow \infty$, the stability region for the finite battery case approaches to that for the infinite battery case.

%

\section{Collision Channel with Probabilistic Erasures}\label{sec:collision}

\begin{figure}[t]
\centering
\epsfig{file=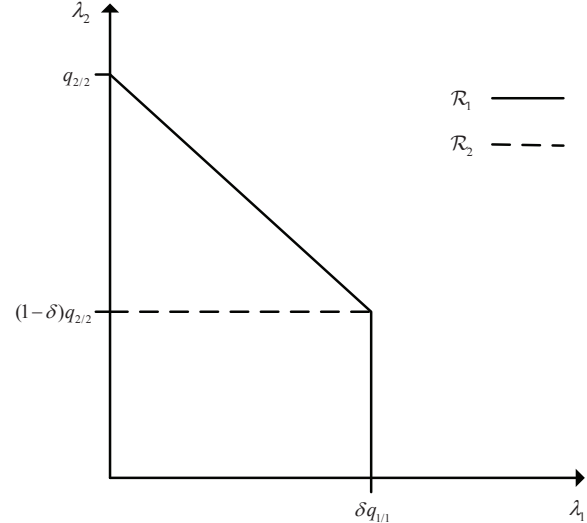,angle=0,width=0.42\textwidth}
\caption{The stability region for the case of collision channel with probabilistic erasures (solid and dotted lines depict the subregions $\mathcal{R}_1$ and $\mathcal{R}_2$, respectively.)}
\label{fig:R_collision}
\end{figure}

For the completeness of our discussion, we present the stability conditions for the collision channel case with probabilistic erasures.
%
%
The stability region is given by:
\begin{equation}
    \mathcal{R} = \mathcal{R}_1 \bigcup \mathcal{R}_2
\end{equation}
where
\begin{equation}
\mathcal{R}_{1}=\left\{ (\lambda_{1},\lambda_{2}): \frac{\lambda_1}{q_{1/1}} + \frac{\lambda_2}{q_{2/2}} < 1 , 0 \leq \lambda_{1} \leq \delta q_{1/1}  \right\}
\end{equation}
and
\begin{equation}
\mathcal{R}_{2}=\left\{ (\lambda_{1},\lambda_{2}): \lambda_1 < \delta q_{1/1} , 0 \leq \lambda_{2} \leq (1-\delta) q_{2/2}  \right\}
\end{equation}
The proof is omitted for brevity.
%
%
It is trivial to observe  that $\mathcal{R}_{2} \subset \mathcal{R}_{1}$ and, thus, $\mathcal{R}  = \mathcal{R}_1$. The optimal $p^*$ achieving the boundaries is always $p^{*}=0$. It is intuitive that the well-designed cognitive access protocol will not allow the secondary node to access the channel when the primary node is transmitting. This is because such simultaneous transmissions would definitely result in a collision. The stability region is depicted in the Fig.~\ref{fig:R_collision}. Since the stability region is identical with the subregion $\mathcal{R}_1$ for the case of $\eta \leq 0$ with multipacket reception capability in Fig. \ref{fig:R1_eta_leq}, the stability region for the collision case is a subset of that for the case with the multipacket reception capability.

\section{Conclusion} \label{sec:conclusion}

We employed an opportunistic multiple access protocol that observes the priorities among the users to better utilize the limited energy resources. Owing to the multipacket reception capability, the secondary node not only utilizes the idle slots but also can take advantage of such an additional reception by transmitting along with the primary node by randomly accessing the channel in a way that does not adversely affect the quality of the communication over the primary link. Consequently, at a given input rate of the primary source, we could choose the optimal access probability by the secondary transmitter to maximize its own throughput and this maximum was also identified. The result is obtained for both cases when the capacity of the battery at the primary node is infinite and also finite. This initial research provides some insights on how to run such a network of nodes having different energy constraints. Extending the approach proposed here to more realistic environments with multiple set of source-destination pairs, although highly desirable, presents serious difficulties due to the interaction between the nodes.

\bibliographystyle{IEEEtran}
\bibliography{bib_cooperation}

\end{document}